# Electromagnetic and Gravitational Waves: the Third Dimension

Gerald E. Marsh

Argonne National Laboratory (Ret)

5433 East View Park

Chicago, IL 60615

E-mail: gemarsh@uchicago.edu

**Abstract.** Plane electromagnetic and gravitational waves interact with particles in such a way as to cause them to oscillate not only in the transverse direction but also along the direction of propagation. The electromagnetic case is usually shown by use of the Hamilton-Jacobi equation and the gravitational by a transformation to a local inertial frame. Here, the covariant Lorentz force equation and the second order equation of geodesic deviation followed by the introduction of a local inertial frame are respectively used. It is often said that there is an analogy between the motion of charged particles in the field of an electromagnetic wave and the motion of test particles in the field of a gravitational wave. This analogy is examined and found to be rather limited. It is also shown that a simple special relativistic relation leads to an integral of the motion, characteristic of plane waves, that is satisfied in both cases.

## Introduction

It has been known for some time that the interaction of a plane electromagnetic wave with a test charge induces a motion, exclusive of that due to radiation pressure,[1] along the direction of propagation.[2] This is usually demonstrated by use of the Hamilton-Jacobi equation. A simpler approach, using the relativistic Lorentz force equation, will be used here to illustrate a class of motions that initially appears very similar to those produced by plane gravitational waves. With regard to the latter, induced motion along the direction of propagation has also been known for some time,[3] although in this case there is greater confusion in the literature since, when sources are present, it is possible to choose gauges[4], such as the Lorentz gauge, where non-radiative parts of the metric obey wave equations. The origin of this confusion dates back to at least Eddington[5], and a very clear exposition of this problem has been given by Flanagan and Hughes.[6]

In both cases it will be seen that the momentum along the direction of propagation is related to the time-like component of the 4-momentum. This is due to the wave nature of the propagation and the relation between the two components of momentum is obtained in the first section that discusses the interaction of an electromagnetic wave with an electron. The second section covers the gravitational case. It shows the significantly different behavior of test particles under the influence of a gravitational wave compared to a charged particle responding to an electromagnetic one.

## Interaction of a plane-polarized electromagnetic wave with an electron

A charged particle under the influence of a continuous plane electromagnetic wave can only gain momentum in the direction of propagation (the behavior under interaction with a short pulse of radiation is, however, more complex[7]). The momenta in the transverse direction will be oscillatory and will not lead to a net momentum gain. As a result, one can expect a relationship to exist between the time-like component of the 4-momentum and the momentum in the direction of propagation.

Much of the behavior of the particle can be understood from its motion in a frame where, as put by Landau & Lifshitz,[2] the particle is "at rest on the average". Although it is the interaction of the particle with the magnetic component of the electromagnetic plane wave that is responsible for the particle's motion in the direction of propagation, it will be seen that one need not include the magnetic component in the equations of motion to determine the momentum in the direction of propagation. This may be found from the time-like component.

The relevant 4-vectors and relations, using the conventions from Jackson,[8] needed to derive the required relationship are

$$x = (\mathbf{x}, ict), \quad k = \left(\mathbf{k}, \frac{i\omega}{c}\right), \quad p = \left(\mathbf{p}, \frac{iE}{c}\right), \quad dt = d\tau, \quad E = \gamma mc^2,$$
(1)

where the symbols have their usual meanings. The equation of motion is

$$\frac{d\mathbf{p}}{dt} = e\left(\mathbf{E} + \frac{\mathbf{v} \times \mathbf{B}}{c}\right).$$
(2)

For a plane polarized wave traveling in the $\hat{\mathbf{n}}$-direction, the following relations hold

$$\mathbf{B} = \hat{\mathbf{n}} \times \mathbf{E} \quad \text{and} \quad \hat{\mathbf{n}} \cdot \mathbf{E} = \hat{\mathbf{n}} \cdot \mathbf{B} = \mathbf{0}.$$
(3)

Thus, using the first of Eqs.(3) in Eq. (2) and expanding the resulting triple product, one obtains

$$\frac{d(\gamma \mathbf{v})}{dt} = \frac{e}{m}\left[\left(1 - \frac{\mathbf{v} \cdot \hat{\mathbf{n}}}{c}\right)\mathbf{E} + \hat{\mathbf{n}}\frac{\mathbf{E} \cdot \mathbf{v}}{c}\right],$$
(4)

where $\mathbf{p} = \gamma m\mathbf{v}$. Dotting through with $\hat{\mathbf{n}}$ yields the simple expression

$$\frac{d(\gamma \mathbf{v} \cdot \hat{\mathbf{n}})}{dt} = \frac{e}{m}\frac{\mathbf{E} \cdot \mathbf{v}}{c}.$$
(5)

Remembering that the 4-velocity and 4-momentum are perpendicular, Eqs (1) imply that

$$\frac{dE}{dt} = \dot{\mathbf{p}} \cdot \mathbf{v},$$
(6)

and Eq. (2), when dotted with $\mathbf{v}$, gives

$$\dot{\mathbf{p}} \cdot \mathbf{v} = e\mathbf{E} \cdot \mathbf{v}.$$
(7)

Thus,

$$\frac{d(\gamma \mathbf{v} \cdot \hat{\mathbf{n}})}{dt} = \frac{1}{mc}\frac{dE}{dt}.$$
(8)

Integrating with respect to time and evaluating the right hand side between the limits given by the initial energy and that at time *t* results in

$$\left(1 - \frac{\mathbf{v} \cdot \hat{\mathbf{n}}}{c}\right) = \frac{E_0}{mc^2}, \quad (9)$$

where $E_0$ is the initial energy of the particle. The right hand side of this equation is a constant and can be written as

$$\frac{E_0}{mc^2} = \frac{1}{mc^2}\left(mc^2 + \frac{1}{2}mv_0^2 + \ldots\right). \quad (10)$$

A similar derivation has been given by Kolomenskii and Lebedev.[9] Equation (10) implies that if the initial velocity $v_0$ vanishes, the right hand side of Eq. (9) reduces to unity. This will be assumed to be the case in what follows.

Now let $\mathbf{v} \cdot \hat{\mathbf{n}} = v_3$ so that the space-time dependence of a plane wave propagating in the $x_3$-direction would be $kx_3 - \omega t$, where $k = \omega/c$. This dependence may be written as

$$\omega\left(\frac{x_3}{c} - t\right) = \phi, \quad (11)$$

which also defines $\phi$. Taking the derivative of $\phi$ with respect to proper time gives

$$\frac{d\phi}{d\tau} = \omega\left(\frac{\dot{x}_3}{c} - 1\right). \quad (12)$$

The right hand side of Eq. (12) is the same as the negative of the left hand side of Eq. (9) when $\mathbf{v} \cdot \hat{\mathbf{n}} = v_3$, so that if the initial velocity vanishes, $d\phi/d\tau = -\omega$. Now, Eq. (12) may be rewritten as

$$\frac{1}{imc}\left(imc - i\omega m\dot{x}_3\right) = \frac{1}{imc}(p_4 - ip_3) = -1, \quad (13)$$

or,

$$(p_4 - ip_3) = -imc. \quad (14)$$

Thus, for a plane wave having the space time dependence of Eq. (11), we have that

$$\frac{dp_3}{d\tau} = -i\frac{dp_4}{d\tau}. \quad (15)$$

Equation (14) is, for the conditions given, a constant of the motion, and Eq. (15) will be used to determine the momentum in the $x_3$-direction in the electromagnetic case, and will be found to also be satisfied in the gravitational case.

Equation (15) is a consequence of the wave depending on the phase $(kx^3 - \omega t)$ rather than being a general function of $x^3$ and $t$. Since the velocity of propagation of a plane monochromatic wave is the velocity with which the planes of constant phase move, taking the derivative with respect to time of $(kx^3 - \omega t) = $ const. gives $dx^3/dt = c$. Multiplying by $m$ and using the definitions in Eqs. (1) gives $p^3 = -ip^4$, and taking the derivative with respect to proper time gives Eq. (15).

The motion of the particle may now be determined by assuming that the vector potential of the plane wave has the form

$$\mathbf{A}(\phi) = \hat{\mathbf{i}}\, \mathcal{A}_1(\phi) + \hat{\mathbf{j}}\, \mathcal{A}_2(\phi), \tag{16}$$

where,

$$\mathcal{A}_1(\phi) = A_0 \sin\phi, \qquad \mathcal{A}_2(\phi) = B_0 \cos\phi. \tag{17}$$

$B_0 = 0$ corresponds to linear polarization, $A_0 = B_0$ to circular polarization, and $A_0 \neq B_0$, where both are not zero, to elliptical polarization.

The following will show that it is only necessary to consider the electric component of the plane wave (as mentioned above). Eq. (2) determines the $x_1$ and $x_2$ components of the force as

$$\frac{dp_1}{d\tau} = -\frac{e}{c}\frac{\partial \mathcal{A}_1(\phi)}{\partial \tau},$$

$$\frac{dp_2}{d\tau} = -\frac{e}{c}\frac{\partial \mathcal{A}_2(\phi)}{\partial \tau}, \tag{18}$$

where $\mathbf{E} = -\frac{1}{c}\frac{\partial \mathbf{A}}{\partial t}$ has been used. From Eqs. (1), (6), and (7)

$$\frac{dp_4}{d\tau} = \frac{i}{c}\frac{dE}{dt} = \frac{i}{c}\frac{d\mathbf{p}}{dt}\cdot\mathbf{v} = \frac{ie}{c}\mathbf{E}\cdot\mathbf{v}, \tag{19}$$

which in turn may be written as

$$\frac{dp_4}{d\tau} = \frac{ie}{mc}\, m\mathbf{v}\cdot\mathbf{E} = \frac{ie}{mc}(E_1 p_1 + E_2 p_2). \tag{20}$$

Using Eq. (16) and noting that $\frac{\partial \phi}{\partial t} = -\omega$, Eq. (20) becomes

$$\frac{dp_4}{d\tau} = -\frac{ie}{mc^2}\left(\frac{A_1(\eta)}{}p_1 + \frac{A_2(\eta)}{}p_2\right). \tag{21}$$

From Eq. (15), the third component of the force is then

$$\frac{dp_3}{d\tau} = -\frac{e}{mc^2}\left(\frac{A_1(\eta)}{}p_1 + \frac{A_2(\eta)}{}p_2\right). \tag{22}$$

Equations (18) may be immediately integrated to

$$p_1 = -\frac{e}{c}A_1(\eta),$$
$$p_2 = -\frac{e}{c}A_2(\eta), \tag{23}$$

so that Eq. (22) becomes

$$\frac{dp_3}{d\tau} = \frac{e^2}{mc^3}\left(\frac{A_1(\eta)}{}A_1(\eta) + \frac{A_2(\eta)}{}A_2(\eta)\right). \tag{24}$$

Remembering that $d\eta/d\tau = -$ this is integrates to

$$p_3 = -\frac{e^2}{2mc^3}\left(A_1(\eta)^2 + A_2(\eta)^2\right). \tag{25}$$

Using Eqs. (17),

$$x_3 = -\frac{e^2}{2m^2c^3}\left(A_0^2\int_0^\eta \sin^2\eta\, d\eta + B_0^2\int_0^\eta \cos^2\eta\, d\eta\right). \tag{26}$$

This is again easily integrated and doing so results in

$$kx_3 = \frac{e^2}{4m^2c^4}\left(\left(A_0^2 + B_0^2\right)\eta + \frac{1}{2}\left(B_0^2 - A_0^2\right)\sin 2\eta\right). \tag{27}$$

The first term in the parentheses of Eq. (27) may be eliminated by a Lorentz transformation to a frame where the particle is "at rest on the average". This is the frame where only the oscillatory motion is evident and will be called here the "rest frame" in quotes. To determine the velocity associated with the transformation, one uses the definition of $\eta$ and takes the derivative of Eq. (27) with respect to *t* using only the first term within the brackets, and solves for the velocity. Its value will play no role in what follows.

The expressions for $p_1$ and $p_2$ in Eq. (23) may be integrated to give

$$kx_1 = -\frac{eA_0}{mc^2}\cos\;,$$

$$kx_2 = \frac{eB_0}{mc^2}\sin\;. \tag{28}$$

Note that if one defines $Q_1 = \dfrac{eA_0}{kmc^2}$ and $Q_2 = \dfrac{eB_0}{kmc^2}$, then Eqs. (28) give the equation of the ellipse

$$\frac{x_1^2}{Q_1^2} + \frac{x_2^2}{Q_2^2} = 1. \tag{29}$$

For a linearly polarized wave, $B_0 = 0$ and a plot of $x_3$ as a function of $x_1$ results in the well known figure eight plot shown in Fig. 1.

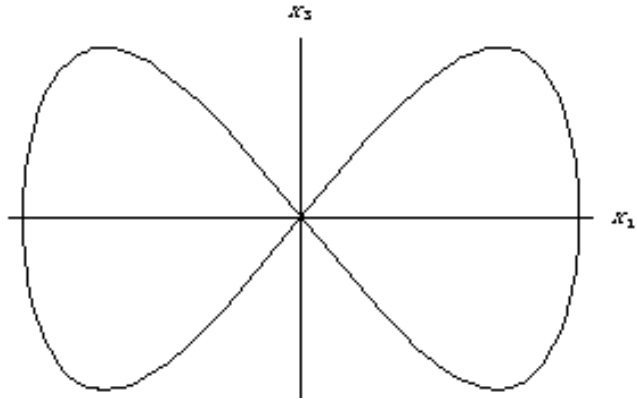

Figure 1. Electron motion in the "rest frame" under influence of a linearly polarized plane electromagnetic wave.

For an elliptically polarized wave, where $A_0$  $B_0$, a parametric plot, where the amplitude of the coefficients is varied while maintaining their ratio, may be made of Eq. (27) and Eqs. (28). The result is a surface of the motion as shown in Fig. 2. Note that because of the sin2   term, the saddle shaped surface has two radial nodal lines.

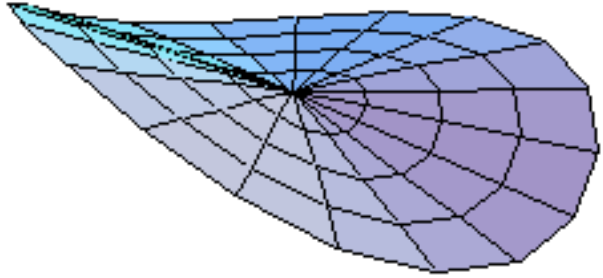

Figure 2. Surface of electron motion in the "rest frame" under influence of an elliptically polarized plane electromagnetic wave.

Increasing radial distance in this figure corresponds increasing wave amplitude while an electron at any given "radius" follows an elliptical path modulated by the sin2  term. This gives a saddle shaped surface of negative curvature. For circular polarization where $A_0 = B_0$, the sin2  term vanishes and the motion is simply circular.

As mentioned above, it is the interaction of the electron with the magnetic component of the plane wave that is responsible for the motion in the $x_3$-direction. Since electric and magnetic components of the wave have the same time dependence given by cos , the equations of motion imply that the velocity gained by the electron due to the electric field will have the time dependence sin . The time dependence of the $v \times B$ term in the equations of motion will then be sin  cos  or sin2 , so that the appearance of this term should be no surprise. It means that the electron can be expected to oscillate with the frequencies  and 2  as has been shown to be the case above.

**Interaction of a plane-polarized gravitational wave with small massive particles**
It will be assumed that the reader is familiar with the equation of geodesic deviation. With reference to Fig. 3, it is given by (Greek indices take the values 0 through 3 and Latin 1 through 3)

$$\frac{D^2 n^\mu}{ds^2} = R^\mu{}_{\phantom{\mu}}\, u\, u\, n\, , \qquad (30)$$

where $D/ds$ is the covariant derivative along a curve. What will be needed here is the second order equation of geodesic deviation. Fundamental work in this area has been done by Bazanski[10] and Kerner[11].

A Taylor expansion of $x^\mu(s, p_1)$ with respect to $p$ is

$$x^\mu(s, p) = x^\mu(s, p_0) + (p - p_0) \left.\frac{x^\mu}{p}\right|_{s, p_0} + \frac{1}{2} (p - p_0)^2 \left.\frac{^2 x^\mu}{p^2}\right|_{s, p_0} + \ldots \tag{31}$$

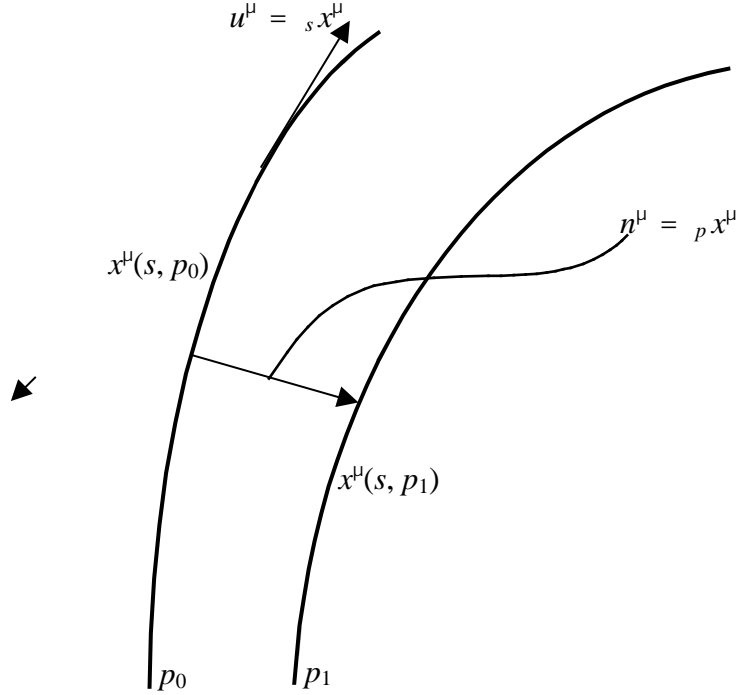

Figure 3. A one parameter set of geodesics $x^\mu(s, p)$, with tangent vector $u^\mu = {}_s x^\mu$ and_deviation vector $n^\mu = {}_p x^\mu$.

The second order term in the expansion of Eq. (31) may be written as

$$\frac{1}{2} (p - p_0)^2 \frac{^2 x^\mu}{p^2} = \frac{1}{2} (p - p_0)^2 \frac{n^\mu}{p}. \tag{32}$$

Let the second order deviation vector be defined as

$$b^\mu = \frac{Dn^\mu}{dp} = \frac{n^\mu}{p} + \Gamma^\mu{}_{\nu\lambda} n^\nu n^\lambda.\tag{33}$$

Setting $p_0 = 0$, and using the fact that $n^\mu = \partial_p x^\mu$, the second and third terms in the expansion in Eq. (31) are

$$p\frac{x^\mu}{p} + \frac{1}{2}p^2\frac{\partial^2 x^\mu}{p^2} = p\, n^\mu + \frac{1}{2}p^2 b^\mu - \frac{1}{2}p^2 \Gamma^\mu{}_{\nu\lambda} n^\nu n^\lambda.\tag{34}$$

With regard to Fig. (3), if the geodesic identified by $p_0 = 0$ corresponds to a local inertial frame so that $u^\mu = (1,0,0,0)$, $\Gamma^\mu{}_{\nu\lambda}$ will vanish along this geodesic. Choose this to be the case. This suggests, following Baskran & Grishchuk[12], that one introduce the vector

$$N^\mu = p\, n^\mu + \frac{1}{2}p^2 b^\mu.\tag{35}$$

The spatial components of $N^\mu$ will then give the position of a nearby particle with respect to the local inertial frame. The covariant derivative of $N^\mu$ along this geodesic is

$$\frac{D^2 N^\mu}{ds^2} = p\frac{D^2 n^\mu}{ds^2} + \frac{1}{2}p^2\frac{D^2 b^\mu}{ds^2}.\tag{36}$$

The first term on the right hand side of Eq. (36) can be found from the first order geodesic equation given by Eq. (30). The second term, involving the second order deviation vector $b^\mu$, has been given by Bazanski[10] as

$$\frac{D^2 b^\mu}{ds^2} = R^\mu{}_{\alpha\beta\gamma}\, u^\alpha u^\beta b^\gamma + (R^\mu{}_{\alpha\beta\gamma;\delta} - R^\mu{}_{\alpha\delta\gamma;\beta})u^\alpha u^\beta n^\gamma n^\delta + 4 R^\mu{}_{\alpha\beta\gamma}\, u^\alpha \frac{Dn^\beta}{ds} n^\gamma.\tag{37}$$

Making the substitutions into Eq. (36) results in

$$\frac{D^2 N^\mu}{ds^2} = R^\mu{}_{\alpha\beta\gamma}\, u^\alpha u^\beta N^\gamma + \frac{1}{2}(R^\mu{}_{\alpha\beta\gamma;\delta} - R^\mu{}_{\alpha\delta\gamma;\beta})u^\alpha u^\beta N^\gamma N^\delta + 2 R^\mu{}_{\alpha\beta\gamma}\, u^\alpha \frac{DN^\beta}{ds} N^\gamma.\tag{38}$$

This is the key equation and will be used in what follows. Of course, substituting $N^\mu$ into this equation from Eq. (35) and gathering terms by order in $p$ yields the first order geodesic equation and Eq. (37).

In what follows, attention will be restricted to weak gravitational waves where the metric may be written in synchronous coordinates (where $g_{0i} = 0$, $g_{00} = -1$, and time lines are geodesics normal to the hypersurfaces $t$ = constant) as

$$ds^2 = (\delta_{ij} + h_{ij})dx^i dx^j - c^2 dt^2.\tag{39}$$

The last term on the right hand side of Eq. (38) is of order $h_{ij}^2$ and, consistent with the weak field approximation where only terms linear in $h_{ij}$ are retained, may be ignored in what follows.

Further simplification of Eq. (38) may be had by recognizing that the introduction of a local inertial frame, as was done above to motivate the vector introduced in Eq. (35), means that all of the covariant derivatives in Eq. (38) can be replaced with ordinary partials, and $D^2/ds^2$ may be replaced by $d^2/c^2dt^2$. Further restricting Eq. (38) to the spatial variations, which henceforth will be of interest, yields

$$\frac{d^2 N^i}{dt^2} = R^i_{0\,0}N^j + \frac{1}{2}\left(R^i_{0\,0,k} - R^i_{jk\,0\,0}\right)N^j N^k. \tag{40}$$

With reference again to Fig. 3, as discussed earlier, $p_0$ defines a local inertial frame satisfying $x^i(t) = 0$, and a point on the geodesic $p_1$ will have a position $x^i_0$ at time $t = 0$. Eq. (40) will be used to find the trajectory of the point $x^i_0$. The deviation vector $N^i$ may then be written as

$$N^i = x^i_0 + \xi^i(t), \tag{41}$$

where, again, $x^i_0$ is the original position at time $t = 0$, and $\xi_i(t)$ is the perturbation caused by the passing gravitational wave in the frame of the inertial coordinates $\xi^i$. Since $x^i_0$ is not a function of time, and since the curvature tensor and $\xi^i$ are of first order in $h_{ij}$, Eq. (40), retaining terms only to first order in $h_{ij}$, becomes

$$\frac{d^2 \xi^i}{dt^2} = R^i_{0\,0}x^j_0 + \frac{1}{2}\left(R^i_{0\,0,k} - R^i_{jk\,0\,0}\right)x^j_0 x^k_0. \tag{42}$$

The surviving components of the curvature tensor for a plane gravitational wave in the transverse traceless gauge have been given by a number of authors including Misner, Thorne, and Wheeler[13]. Cosideration in what follows will be restricted to the + polarization. With reference to the metric of Eq. (39), the curvature tensor components all have the form $\frac{1}{2}\ddot{h}_{ij}$, where the dot corresponds to differentiation with respect to $t$. Equation (42) becomes, using conventions that conform to the literature cited,

$$\frac{d^2 \xi^i}{dt^2} = -\frac{1}{2}\ddot{h}^i_j x^j_0 - \frac{1}{2}\left(\ddot{h}^i_{j,k} - \frac{1}{2}\,{}^i_m\ddot{h}_{kj,m}\right)x^j_0 x^k_0. \tag{43}$$

The derivation of the term $-\frac{1}{2}\ddot{h}^i_{j,k}x_0^j x_0^k$ requires the use of the Bianchi identity contracted with the symmetric product $x_0^j x_0^k$ to show the symmetry of $R^i_{0 0 j,k}$ with respect to j and k, which results in the factor of _ in this term rather than the _ that might be anticipated. The term $\frac{1}{4}\,{}_m\ddot{h}^i_{kj,m}x_0^j x_0^k$ is arrived at as follows: $h^i_j$ has the form

$$h^i_j = h_+ p^i_j \sin(kx^3 - \omega t), \tag{44}$$

where $p^i_j$ is the polarization tensor, whose components here are restricted to $p^1_1 = 1$, and $p^2_2 = -1$. From the form of Eq. (44) it is readily seen that taking the derivative of $h^i_j$ with respect to $x^0$ is equivalent to applying the operator $-\eta^i_m \partial_{x^m}$. Because Eq.(44) only depends on $x^3$, only $\partial_{x^3}$ is non-zero.

Extending Eq. (43) to include the $\xi^0$-term, which will play only a very limited role in what follows, results in $\frac{d^2 \xi^0}{dt^2} = -\frac{1}{4}\ddot{h}_{kj,0}x_0^k x_0^j$. Using this and substituting Eq. (44) into Eq. (43) results in the following set of equations:

$$\frac{d^2\xi^0}{dt^2} = \frac{1}{4}h_+ k\omega^2\left(x_0^{2^2} - x_0^{1^2}\right)\cos(kx^3 - \omega t),$$

$$\frac{d^2\xi^1}{dt^2} = \frac{1}{2}x_0^1 h_+\omega^2\sin(kx^3 - \omega t) + \frac{1}{2}h_+ x_0^1 x_0^3 k\omega^2\cos(kx^3 - \omega t),$$

$$\frac{d^2\xi^2}{dt^2} = -\frac{1}{2}x_0^2 h_+\omega^2\sin(kx^3 - \omega t) - \frac{1}{2}h_+ x_0^2 x_0^3 k\omega^2\cos(kx^3 - \omega t),$$

$$\frac{d^2\xi^3}{dt^2} = \frac{1}{4}h_+ k\omega^2\left(x_0^{2^2} - x_0^{1^2}\right)\cos(kx^3 - \omega t). \tag{45}$$

Comparing the first and last of these equations, shows that

$$\frac{d^2\xi^0}{dt^2} = \frac{d^2\xi^3}{dt^2}. \tag{46}$$

The variations of the position and energy of particles in the $\xi$-frame, where the particles are on the average at rest, are small and proportional to $h_+$. As a result, Eq. (46) can be put into the same form as Eq. (15).

Equations (45) are easily integrated, and with appropriate constants of integration yield

$$\begin{aligned}
\xi^0 &= x_0^0 - \frac{1}{4}h_+ k\left(x_0^{2^2} - x_0^{1^2}\right)\cos(kx^3 - \omega t),\\
\xi^1 &= x_0^1 - \frac{1}{2}x_0^1 h_+ \sin(kx^3 - \omega t) - \frac{1}{2}h_+ x_0^1 x_0^3 k \cos(kx^3 - \omega t),\\
\xi^2 &= x_0^2 + \frac{1}{2}x_0^2 h_+ \sin(kx^3 - \omega t) + \frac{1}{2}h_+ x_0^2 x_0^3 k \cos(kx^3 - \omega t),\\
\xi^3 &= x_0^3 - \frac{1}{4}h_+ k\left(x_0^{2^2} - x_0^{1^2}\right)\cos(kx^3 - \omega t).
\end{aligned} \qquad (47)$$

These are essentially the same equations as those found by Grishchuk[14] and, when $x_0^\mu$ is set equal to $x^\mu$, correspond to a coordinate transformation between the local inertial and synchronous reference frames.  Henceforth, $\xi^0$ will play no role.

The effect of the wave, represented by Eqs. (47), can be seen in 3-dimensional space as follows: consider the motion in the $\xi^1$, $\xi^2$-plane, and set $x_0^3 = 0$. Introduce a circle of test masses by transforming the initial positions $x_0^1, x_0^2$ to the cylindrical coordinates $x_0^1 = r_0 \cos\phi$, $x_0^2 = r_0 \sin\phi$. $r_0$ will be held fixed and $\phi$ will be allowed to vary so as to show the effect of the wave on the motion of a set of test masses distributed around the central geodesic identified above as $p_0$. Without including motion in the $\xi^3$-direction, the effect on such a circle of test masses is shown in Fig. 4(a).  The vertical direction corresponds to the change in the phase $(kx^3 - \omega t)$ and is equivalent to the time evolution. The vertical lines in the grid outlining the figure would correspond to the trajectory of a set of small test masses.  Horizontal cross sections of constant phase display the motion usually depicted in textbooks of the effect of a plane gravitational wave on a ring of test masses.

Figure 4(b) includes the (exaggerated) motion in the $\xi^3$-direction. If one considered the effect of the wave on a disk of test masses, horizontal cross sections of constant phase of this figure would look like the surface associated with electron motion in the electromagnetic case shown in Fig. 2. The horizontal grid lines of Fig. 4(b) correspond to the edge of this surface.

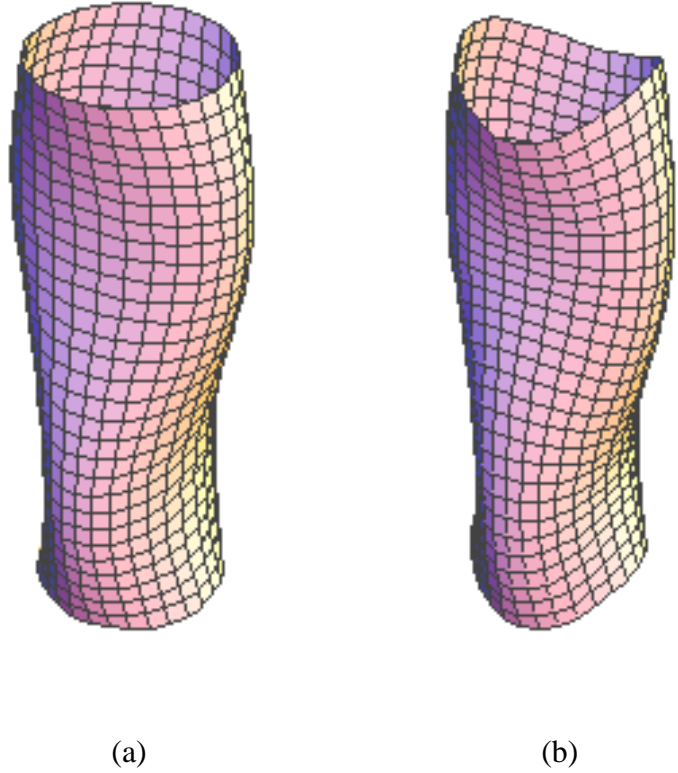

(a)                  (b)

Figure 4. The vertical axis corresponds to the time evolution of the phase $(kx^3 - \omega t)$: (a) shows the time evolution in the $x^1, x^2$-plane ignoring any motion in the $x^3$-direction; (b) shows the motion with that in the $x^3$-direction included.

Unlike the motion in Fig. 2, however, the test particles of Fig. 4(b) do not follow an elliptical path around the central trajectory of the wave modulated by a trigonometric function of a double angle. This is also the case for a circularly or elliptically polarized gravitational wave.

Instead, the motion of an individual test mass located at $\left(x_0^1, x_0^2\right)$, which is not on one of the radial node lines of the constant phase surface of Fig. 4(b), is a combination of the motion in the $x^1, x^2$-plane with that in the $x^3$-direction. This is shown in Fig. 5.

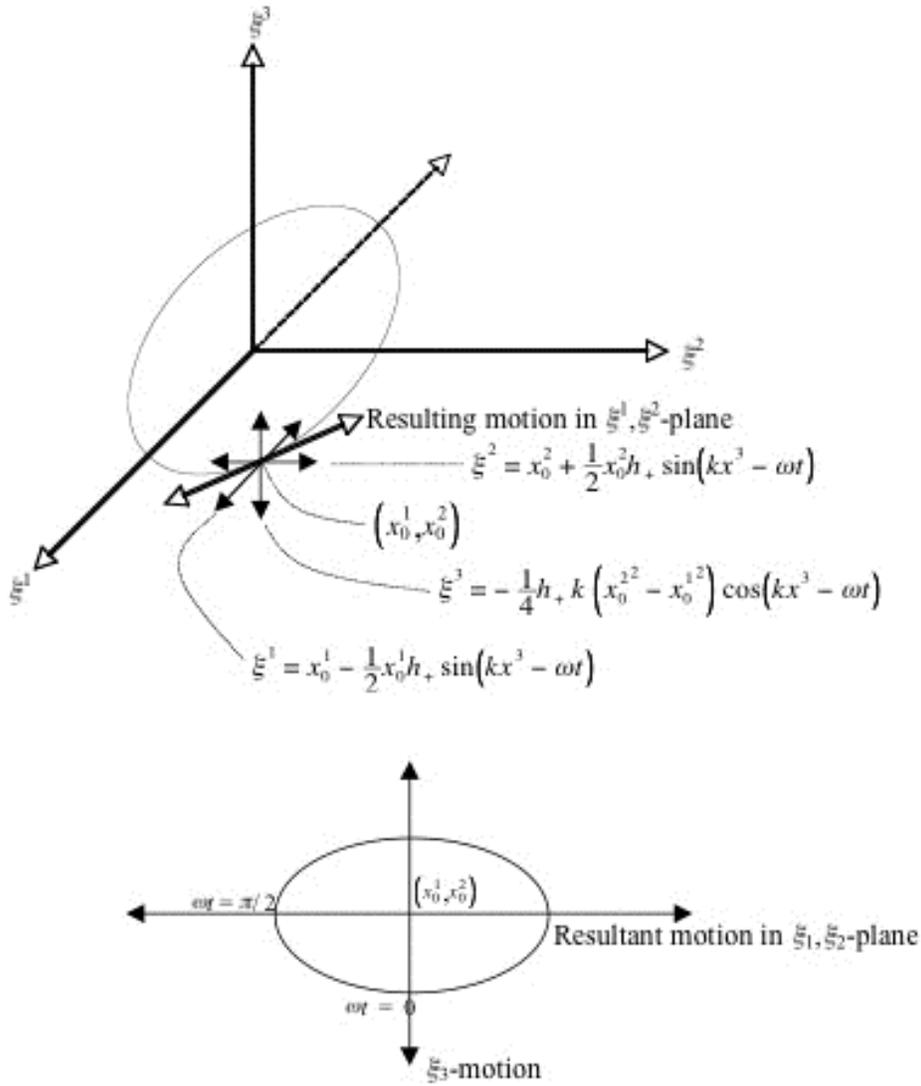

Figure 5. (a) shows the components of the motion. The ellipse lies in the $\xi^1, \xi^2$-plane, and corresponds to a horizontal constant phase section of Fig. 4(a). (b) shows the elliptical motion of a test mass located at $(x_0^1, x_0^2)$. Note that the plane of the ellipse in Fig. 5(b) is perpendicular to the $\xi^1, \xi^2$-plane.

The surfaces of constant phase in Fig. 4(b) appear to have negative curvature like Fig. 2. That these phase surfaces do indeed have negative curvature can be seen as follows: In the local inertial coordinates of Eqs. (47), Fig. 4(b) was constructed by setting $x_0^3 = 0$ and then introducing the cylindrical coordinates $x_0^1 = r_0 \cos\phi$, $x_0^2 = r_0 \sin\phi$. The resulting coordinates are

$$\xi_1 = r_0 \cos\phi - \frac{1}{2} h_+ r_0 \sin\phi \cos\varphi,$$
$$\xi_2 = r_0 \sin\phi + \frac{1}{2} h_+ r_0 \sin\phi \sin\varphi,$$
$$\xi_3 = \frac{1}{4} r_0^2 h_+ k \cos\phi \cos 2\varphi, \qquad (48)$$

where the phase $\varphi = (kx_3 - \omega t)$, when set equal to a constant, results in the constant phase surfaces of Fig. 4(b). From Eqs. (48), the spatial metric in local inertial coordinates for a fixed value of $r_0$ is

$$ds^2 = d\xi_1^2 + d\xi_2^2 + d\xi_3^2$$
$$= r_0^2 d\phi^2 \left[ \left(1 - \frac{1}{2} h_+ \sin\varphi \right)^2 \sin^2\phi + \left(1 + \frac{1}{2} h_+ \sin\varphi \right)^2 \cos^2\phi + \frac{1}{4} r_0^2 h_+^2 k^2 \cos^2\phi \sin^2 2\varphi \right]. \qquad (49)$$

The spatial curvature of the constant phase surface can be determined by computing the difference between the circumference of a circle of fixed radius $r_0$ in Euclidean space and comparing it to one in the local inertial frame; that is,

$$2\pi r_0 - \int_0^{2\pi} \left[ \left(1 - \frac{1}{2} h_+ \sin\varphi \right)^2 \sin^2\phi + \left(1 + \frac{1}{2} h_+ \sin\varphi \right)^2 \cos^2\phi + \frac{1}{2} r_0^2 h_+^2 k^2 \cos^2\phi \sin^2 2\varphi \right]^{\frac{1}{2}} r_0 d\phi.$$

(50)

When evaluating this expression for the spatial curvature, it is important to recall the limits of the linear approximation. The approach of using the equation of geodesic deviation to determine the spatial variations is only valid if the magnitude of the deviation vector is small compared to the length (often called the inhomogeneity scale) over which the Riemann tensor changes. In particular, the magnitude of $r_0$ must be much less than the wavelength $\lambda$ of the gravitational wave.

The difference in Eq. (50) vanishes for $r_0 = 0$, and is always negative for $r_0 > 0$, consistent with the apparent negative curvature of the constant phase surfaces seen in Fig. 4(b). When plotting the difference given by Eq. (50) for different values of $r_0$, one obtains a linear decrease for $r_0 \ll \lambda$, the wavelength of the gravitational wave. The difference becomes non-linear as $r_0$ approaches . This non-linear behavior is due to exceeding the range over which the linear approximation leading to the metric of Eq. (39) is valid.

It should be emphasized that this should not be interpreted as meaning that the spatial curvature of a space-like hypersurface, where *t* equals a constant, is negative. A gravitational wave carries positive energy, which results in a positive space-time curvature.

**Comparing electromagnetic and gravitational waves**

There is indeed an analogy between electromagnetic and gravitational waves. Both have two linear polarizations that may be combined to yield circular or elliptically polarized waves. But their effect on a set of test particles is very different.

The dynamics of a charged particle is due entirely to the Lorentz force. The longitudinal motion under the influence of a linearly or elliptically polarized continuous plane electromagnetic wave oscillates with a frequency twice that of the transverse frequency. This is shown by Eq. (27) and Figs. (1) and (2) above. The longitudinal motion vanishes in the case of a circularly polarized wave.

The motion of a set of test particles under the influence of a plane gravitational wave differs considerably from the electromagnetic case. Yet, there are similarities: not only do both have two independent polarization states, but when one includes the longitudinal motion, the surface associated with the motion of a charged particle responding to an elliptically polarized wave (Fig. (2)) is similar to the constant phase surfaces of a set of particles driven by a plane gravitational wave (Fig. 4(b)); in both cases the latter surfaces derive their longitudinal motion from trigonometric double angle functions. But in the gravitational case, the test particles do not move around the central geodesic. Instead, they have an oscillatory motion in the transverse plane, which when coupled to the longitudinal motion, leads to the particles moving in ellipses whose planes are perpendicular to the transverse plane (Fig. (5)).

If one were to include the $h_\times$ polarization, the $\xi_3$-motion in Eq. (48) would have the additional term

$$\frac{1}{4} k\, h_\times r_0^2 \sin 2\phi \sin \psi. \tag{51}$$

The constant phase surfaces would still have the appearance shown in Fig. 4(b), but as $\psi$ advanced from 0 to $2\pi$, the surfaces would rotate about the vertical direction in the figure. The sin $\psi$ and cos $\psi$ terms combine with the double angle terms in a counter-intuitive way,[15] such that a change in phase of $\psi$ corresponds to a rotation about the direction of propagation by $\psi/2$. For circular polarization, where $h_+ = \pm i\, h_\times$, the longitudinal motion does not vanish for all $\psi$ in contrast to the electromagnetic case.

**Footnotes**

[1] For a discussion of radiation pressure and its relation to radiation damping see: K. Hagenbuch, "Free electron motion in a plane electromagnetic wave," *Am. J. Phys*. **45**, 693-696 (1977).

[2] L.D. Landau and E.M. Lifshitz, *The Classical Theory of Fields* (Pergamon Press, Oxford 1962), p. 128 [Modified in the 4$^{th}$ edition] and p. 134.

[3] R. Adler, M. Bazin, and M. Schiffer, *Introduction to General Relativity* (McGraw-Hill, New York 1965), §8.5.

[4] In general relativity, gauge transformations are coordinate transformations.

[5] A.S. Eddington, *The Mathematical Theory of Relativity* (Cambridge University Press, London 1960), §57.

[6] E. E Flanagan and S.A. Hughes, "The basics of gravitational wave theory" *New Journal of Physics* **7**, 204 (2005).

[7] See, for example: A.L. Galkin, *et al*., "Dynamics of an electron driven by relativistically intense laser radiation", *Phys. Plasmas* **15**, 023104 (2008).

[8] J.D. Jackson, *Classical Electrodynamics* (John Wiley & Sons, New York 1962).

[9] A.A. Kolomenskii and A.N. Levedev, "Self-Resonant Particle Motion in a Plane Electromagnetic Wave", *Sov. Phys.-Doklady* **7**, 745-747 (1963).

[10] S.L. Bazanski, "Kinematics of relative motion of test particles in general relativity", *Ann. Inst. H. Poincare* **A27**, 115-144 (1977).

[11] R. Kerner, J.W. van Holten, and R. Colistete Jr, "Relativistic Epicycles: another approach to geodesic deviations", *Class. Quant. Grav.* **18**, 4725-4742 (2001); R. Kerner and S. Vitale, *Proc. of Science—5$^{th}$ Int. School on Field Theory and Gravitation* (2009).

[12] D. Baskaran and L.P. Grishchuk, "Components of the gravitational force in the field of a gravitational wave", *Class. Quant. Grav.* **21**, 4041-4062 (2004).

[13] C.W. Misner, K.S. Thorne, and J.A. Wheeler, *Gravitation* (W.H. Freeman & Co., San Francisco 1973).

[14] L.P. Grishchuk, "Gravitational waves in the cosmos and the laboratory", *Sov. Phys. Usp*. **20**, 319-334 (1977). Grishchuk considered a wave traveling in the $x^1$-direction. His

equations (12) or (13) may be transformed into those used here by the transformation $x^1 \to x^3, x^2 \to x^1, x^3 \to x^2$.

[15] C.W. Misner, K.S. Thorne, and J.A. Wheeler, *op. cit.*, §37.2.